\PassOptionsToPackage{hyphens}{url}
\PassOptionsToPackage{dvipsnames,svgnames,x11names}{xcolor}
\documentclass[
  12pt,
  letterpaper,
  DIV=11,
  numbers=noendperiod]{scrartcl}
\usepackage{xcolor}
\usepackage{amsmath,amssymb}
\usepackage{iftex}
  \usepackage[T1]{fontenc}
  \usepackage[utf8]{inputenc}
  \usepackage{textcomp} 
\usepackage{lmodern}
\ifPDFTeX\else
\fi
\IfFileExists{upquote.sty}{\usepackage{upquote}}{}
\IfFileExists{microtype.sty}{
  \usepackage[]{microtype}
  \UseMicrotypeSet[protrusion]{basicmath} 
}{}
\makeatletter
\@ifundefined{KOMAClassName}{
  \IfFileExists{parskip.sty}{%
    \usepackage{parskip}
  }{
    \setlength{\parindent}{0pt}
    \setlength{\parskip}{6pt plus 2pt minus 1pt}}
}{
  \KOMAoptions{parskip=half}}
\makeatother
\makeatletter
\ifx\paragraph\undefined\else
  \let\oldparagraph\paragraph
  \renewcommand{\paragraph}{
    \@ifstar
      \xxxParagraphStar
      \xxxParagraphNoStar
  }
  \newcommand{\xxxParagraphStar}[1]{\oldparagraph*{#1}\mbox{}}
  \newcommand{\xxxParagraphNoStar}[1]{\oldparagraph{#1}\mbox{}}
\fi
\ifx\subparagraph\undefined\else
  \let\oldsubparagraph\subparagraph
  \renewcommand{\subparagraph}{
    \@ifstar
      \xxxSubParagraphStar
      \xxxSubParagraphNoStar
  }
  \newcommand{\xxxSubParagraphStar}[1]{\oldsubparagraph*{#1}\mbox{}}
  \newcommand{\xxxSubParagraphNoStar}[1]{\oldsubparagraph{#1}\mbox{}}
\fi
\makeatother

\usepackage{longtable,booktabs,array}
\usepackage{calc} 
\usepackage{etoolbox}
\makeatletter
\patchcmd\longtable{\par}{\if@noskipsec\mbox{}\fi\par}{}{}
\makeatother
\IfFileExists{footnotehyper.sty}{\usepackage{footnotehyper}}{\usepackage{footnote}}
\makesavenoteenv{longtable}
\usepackage{graphicx}
\makeatletter
\newsavebox\pandoc@box
\newcommand*\pandocbounded[1]{
  \sbox\pandoc@box{#1}%
  \Gscale@div\@tempa{\textheight}{\dimexpr\ht\pandoc@box+\dp\pandoc@box\relax}%
  \Gscale@div\@tempb{\linewidth}{\wd\pandoc@box}%
  \ifdim\@tempb\p@<\@tempa\p@\let\@tempa\@tempb\fi
  \ifdim\@tempa\p@<\p@\scalebox{\@tempa}{\usebox\pandoc@box}%
  \else\usebox{\pandoc@box}%
  \fi%
}
\def\fps@figure{htbp}
\makeatother

\NewDocumentCommand\citeproctext{}{}
\NewDocumentCommand\citeproc{mm}{%
  \begingroup\def\citeproctext{#2}\cite{#1}\endgroup}
\makeatletter
 \let\@cite@ofmt\@firstofone
 \def\@biblabel#1{}
 \def\@cite#1#2{{#1\if@tempswa , #2\fi}}
\makeatother
\newlength{\cslhangindent}
\newlength{\csllabelwidth}
\newenvironment{CSLReferences}[2] 
 {\begin{list}{}{%
  \setlength{\itemindent}{0pt}
  \setlength{\leftmargin}{0pt}
  \setlength{\parsep}{0pt}
  \ifodd #1
   \setlength{\leftmargin}{\cslhangindent}
   \setlength{\itemindent}{-1\cslhangindent}
  \fi
  \setlength{\itemsep}{#2\baselineskip}}}
 {\end{list}}
\usepackage{calc}

\providecommand{\tightlist}{%
  \setlength{\itemsep}{0pt}\setlength{\parskip}{0pt}}

\usepackage{tcolorbox}
\usepackage{amssymb}
\usepackage{yfonts}
\usepackage{bm}

\newtcolorbox{greybox}{
  colback=white,
  colframe=blue,
  coltext=black,
  boxsep=5pt,
  arc=4pt}
  
\newcommand{\sectionbreak}{\clearpage}

\KOMAoption{captions}{tableheading}
\makeatletter
\@ifpackageloaded{caption}{}{\usepackage{caption}}
\AtBeginDocument{%
\ifdefined\contentsname
  \renewcommand*\contentsname{Table of contents}
\else
  \newcommand\contentsname{Table of contents}
\fi
\ifdefined\listfigurename
  \renewcommand*\listfigurename{List of Figures}
\else
  \newcommand\listfigurename{List of Figures}
\fi
\ifdefined\listtablename
  \renewcommand*\listtablename{List of Tables}
\else
  \newcommand\listtablename{List of Tables}
\fi
\ifdefined\figurename
  \renewcommand*\figurename{Figure}
\else
  \newcommand\figurename{Figure}
\fi
\ifdefined\tablename
  \renewcommand*\tablename{Table}
\else
  \newcommand\tablename{Table}
\fi
}
\@ifpackageloaded{float}{}{\usepackage{float}}
\floatstyle{ruled}
\@ifundefined{c@chapter}{\newfloat{codelisting}{h}{lop}}{\newfloat{codelisting}{h}{lop}[chapter]}
\floatname{codelisting}{Listing}

\makeatother
\makeatletter
\@ifpackageloaded{caption}{}{\usepackage{caption}}
\@ifpackageloaded{subcaption}{}{\usepackage{subcaption}}
\makeatother
\usepackage{bookmark}
\IfFileExists{xurl.sty}{\usepackage{xurl}}{} 
\hypersetup{
  pdftitle={Differentiating Generalized Eigenvalues and Eigenvectors},
  pdfauthor={Jan de Leeuw},
  colorlinks=true,
  linkcolor={blue},
  filecolor={Maroon},
  citecolor={Blue},
  urlcolor={Blue},
  pdfcreator={LaTeX via pandoc}}

\title{Differentiating Generalized Eigenvalues and Eigenvectors}
\author{Jan de Leeuw}
\date{August 12, 2025}
\begin{document}
\maketitle
\begin{abstract}
We give formulae for first and second derivatives of generalized
eigenvalues/eigenvectors of symmetric matrices and generalized singular
values/singular vectors of rectangular matrices when the matrices are
linear or nonlinear functions of a vector of parameters. In addition we
provide functions in R to compute these derivatives, both in the general
case and in various special cases. Formulae are checked against
Jacobians and Hessians computed by numerical differentiation. Some
applications to multivariate data analysis are discussed.
\end{abstract}

\renewcommand*\contentsname{Table of contents}
{
\hypersetup{linkcolor=}
\setcounter{tocdepth}{3}
\tableofcontents
}

\textbf{Note:} This is a working manuscript which will be
expanded/updated frequently. All suggestions for improvement are
welcome. All Rmd, tex, html, pdf, R, and C files are in the public
domain. Attribution will be appreciated, but is not required. The files
can be found at
\url{https://github.com/deleeuw/generalized-eigen-derivatives}.

\sectionbreak

\section{Introduction}\label{introduction}

The generalized eigenvalue decomposition (GEVD) problem for a pair
\((A,B)\) of square symmetric matrices of order \(n\) is to find the
solution \((X,\Lambda)\) of the system \begin{subequations}
\begin{align}
AX&=BX\Lambda,\label{eq-gevdef1}\\
X'BX&=I\label{eq-gevdef2}.
\end{align} 
\end{subequations} We assume that \(B\) is positive definite. The matrix
\(X\) of eigenvectors and the diagonal matrix \(\Lambda\) of eigenvalues
are both square of order \(n\). As an aside we mention that the
generalized eigenvalues of \eqref{eq-gevdef1} and \eqref{eq-gevdef2} are
also the eigenvalues of the asymmetric matrix \(B^{-1}A\) and of the
symmetric matrix \(\smash{B^{-1/2}AB^{-1/2}}\).

GEVD is at the basis of a great many computations in multivariate
statistics. In multinormal small sample theory the derivatives are
needed for the Jacobians in change-of-variable calculations to find the
distribution of many interesting statistics (mathai). The applications
we have in mind, however, are in large-sample statistics, where the
derivatives are used in Delta Method computations of bias estimates,
standard errors, and confidence intervals.

The equations \eqref{eq-gevdef1} and \eqref{eq-gevdef2} implicitly
define \((X(),\Lambda())\) as a function of \((A,B)\). It turns out that
under suitable assumptions on the matrix arguments these implicit
functions are actually differentiable, and this makes it interesting to
compute their derivatives.

There is a humongous and scattered literature on formulae for and
computations of derivatives of eigenvalues and eigenvectors in linear
algebra, numerical mathematics, engineering, multivariate statistics,
and even in physics. Reviewing and collecting all relevant literature is
an impossible task. We only give the basic mathematical references,
which provide the foundations upon which our results will be built (Kato
(\citeproc{ref-kato_84}{1984}), Baumgärter
(\citeproc{ref-baumgartel_85}{1985}), Brustad
(\citeproc{ref-brustad_19}{2019})).

We have written code in R (R Core Team
(\citeproc{ref-r_core_team_25}{2025})) to implement our final formulae
for the first and second derivatives. Code is available at
\url{https://github.com/deleeuw/generalized-eigen-derivatives}. In
addition we have also written R code, available in the same repository,
to numerically verify the final formulae. This uses the functions
jacobian() and hessian() from the excellent R package numDeriv (Gilbert
and Varadhan (\citeproc{ref-gilbert_varadhan_19}{2019})). Of course
numerical verification, especially in the nonlinear case, is limited to
a small number of examples, but if we get the same results as from our
formulae we do gain some confidence.

I should perhaps also mention a previous version of this paper (De Leeuw
(\citeproc{ref-deleeuw_R_07c}{2007})), which has some errors and is not
as complete as the current version.

\sectionbreak

\section{Basic Results}\label{basic-results}

Suppose \(A()\) and \(B()\) are differentiable symmetric matrix valued
functions of order \(n\) on an open subset \(\Theta\) of
\(\mathbb{R}^p\), and suppose at \(\theta_0\in\Theta\) the matrix
\(B(\theta_0)\) is positive definite and the generalized eigenvalues of
\((A(\theta_0),B(\theta_0))\) are all different. Then the ordered
eigenvalues \(\Lambda()\) and the eigenvectors \(X()\) are analytic
functions of the parameters in a neighborhood of \(\theta_0\).

We use subscripts \(i,j=1,\cdots, n\) for the rows and columns of \(A\)
and \(B\), subscripts \(\eta,\nu=1,\cdots,n\) for the eigenvalues and
eigenvectors, and \(s,t=1,\cdots,p\) for the parameters in \(\theta\).
Thus \(x_{i\nu}\) is element \(i\) of eigenvector \(x_\nu\). The partial
derivative of a function \(A()\) on \(\Theta\) with respect to
\(\theta_s\), evaluated at \(\theta_0\), is written as
\(\mathcal{D}_sA(\theta_0)\). In order not to clutter our formulae the
parameter vector where the derivative is evaluated is usually not
explicitly specified. Note also that it is sometimes necessary to use
parentheses to distinguish \((\mathcal{D}_sA)x\) from
\(\mathcal{D}_s(Ax)\).

\subsection{First Partials}\label{first-partials}

Differentiate \eqref{eq-gevdef1} with respect to \(\theta_s\). Then
\begin{equation}
(\mathcal{D}_sA)X+A(\mathcal{D}_sX)=BX(\mathcal{D}_s\Lambda)+
B(\mathcal{D}_sX)\Lambda+(\mathcal{D}_sB)X\Lambda.\label{eq-firstder}
\end{equation} Premultiplying \eqref{eq-firstder} by \(X'\) and
rearranging gives \begin{equation}
\mathcal{D}_s\Lambda=\{X'(\mathcal{D}_sA)X-X'(\mathcal{D}_sB)X\Lambda\}+\{\Lambda X'B(\mathcal{D}_sX)-X'B(\mathcal{D}_sX)\Lambda\}.\label{eq-labfull}
\end{equation} The matrix
\(\Lambda X'B(\mathcal{D}_sX)-X'B(\mathcal{D}_sX)\Lambda\) is
anti-symmetric and consequently has a zero diagonal. Taking the
diagonal\footnote{The diagonal $\text{diag}(X)$ of a square matrix $X$ is a diagonal matrix with the same diagonal as $X$.}
on both sides of \eqref{eq-labfull} gives \begin{equation}
\mathcal{D}_s\Lambda=\text{diag}\{X'(\mathcal{D}_sA)X-X'(\mathcal{D}_sB)X\Lambda\},\label{eq-labsol}
\end{equation} or, for a single eigenvalue \(\lambda_\nu\) with
corresponding eigenvector \(x_\nu\), \begin{equation}
\mathcal{D}_s\lambda_\nu=x_\nu'(\mathcal{D}_sA-\lambda_\nu\mathcal{D}_sB)x_\nu.\label{eq-singval}
\end{equation}

Taking the
off-diagonal\footnote{The off-diagonal $\text{off}(X)$ of a square matrix $X$ is $X$ with its diagonal replaced by zeroes.}
on both sides of \eqref{eq-labfull} gives \begin{equation}
\text{off}\{\Lambda X'B(\mathcal{D}_sX)-X'B(\mathcal{D}_sX)\Lambda\}=-\text{off}\{X'(\mathcal{D}_sA)X-X'(\mathcal{D}_sB)X\Lambda\}.\label{eq-offd}
\end{equation} \(X\) is non-singular, and thus there is a unique square
\(H_s\) such that \(\mathcal{D}_sX=XH_s\). Using this substitution
\eqref{eq-offd} becomes \begin{equation}
\text{off}\{\Lambda H_s-H_s\Lambda\}=-\text{off}\{X'(\mathcal{D}_sA)X-X'(\mathcal{D}_sB)X\Lambda\}.
\end{equation} Switch to subscript notation and solve for \(H_s\). For
\(\nu\neq\eta\) \begin{equation}
\{H_s\}_{\nu\eta}=-\frac{x_\nu'\{\mathcal{D}_sA-\lambda_s\mathcal{D}_sB\}x_\eta}{\lambda_\nu-\lambda_\eta}.\label{eq-offh}
\end{equation} This does not give a value for the diagonal of \(H_s\).
Differentiating \eqref{eq-gevdef2} gives \begin{equation}
X'B(\mathcal{D}_sX)+(\mathcal{D}_sX)'BX+X'\mathcal{D}_sBX=0.\label{eq-fromtwo}
\end{equation} Using \(\mathcal{D}_sX=XH_s\) and taking the diagonal of
\eqref{eq-fromtwo} gives \begin{equation}
h_{\nu\nu}=-\tfrac12 x_\nu'(\mathcal{D}_sB)x_\nu. \label{eq-diagh}
\end{equation} Combining \eqref{eq-offh} and \eqref{eq-diagh} shows that
for the eigenvector corresponding with \(\lambda_\nu\) we have
\begin{equation}
\mathcal{D}_sx_\nu=-\sum_{\substack{\eta=1\\\eta\neq\nu}}^n\frac{x_\eta'\{\mathcal{D}_sA-\lambda_\nu\mathcal{D}_sB\}x_\nu}{\lambda_\eta-\lambda_\nu}x_\eta-\tfrac12 x_\nu'(\mathcal{D}_sB)x_\nu\cdot x_\nu,\label{eq-secder1}
\end{equation} which can also be written as \begin{equation}
\mathcal{D}_sx_\nu=-\sum_{\substack{\eta=1\\\eta\neq\nu}}^n\left\{\frac{x_\eta x_\eta'}{\lambda_\eta-\lambda_\nu}\right\}(\mathcal{D}_sA-\lambda_\nu\mathcal{D}_sB)x_\nu-\tfrac12 (x_\nu'(\mathcal{D}_sB)x_\nu)x_\nu.\label{eq-secder2}
\end{equation}

It is notationally convenient to have a matrix expression for this
derivative. Define the matrices \begin{equation}
(A-\lambda_\nu B)^-:=
\sum_{\substack{\eta=1\\\eta\neq\nu}}^n\left\{\frac{x_\eta x_\eta'}{\lambda_\eta-\lambda_\nu}\right\}=
X(\Lambda-\lambda_\nu I)^+X',\label{eq-geninv}
\end{equation} where \((\Lambda-\lambda_\nu I)^+\) is the Moore-Penrose
inverse of \(\Lambda-\lambda_\nu I\). Matrix \eqref{eq-geninv} is a
reflexive g-inverse of \(A-\lambda_\nu B\) (Rao and Mitra
(\citeproc{ref-rao_mitra_71}{1971}), section 2.5). Of the four Penrose
conditions only the first two are satisfied. To verify this we use
definition \eqref{eq-geninv} and
\(A-\lambda_\nu B=X^{-T}(\Lambda-\lambda_\nu I)X^{-1}\). We find
\begin{subequations}
\begin{align}
(A-\lambda_\nu B)(A-\lambda_\nu B)^-(A-\lambda_\nu B)&=A-\lambda_\nu B,\\
(A-\lambda_\nu B)^-(A-\lambda_\nu B)(A-\lambda_\nu B)^-&=(A-\lambda_\nu B)^-,\\
(A-\lambda_\nu B)(A-\lambda_\nu B)^-&=X^{-T}(I-e_\nu e_\nu')X',\\
(A-\lambda_\nu B)^-(A-\lambda_\nu B)&=X(I-e_\nu e_\nu')X^{-1}.
\end{align}
\end{subequations} We see that \((A-\lambda_\nu B)^-\) is a
Moore-Penrose inverse of \(A-\lambda_\nu B\) if and only if \(B\) is
identically equal to one (and thus \(X'X=XX'=I\)).

Using \eqref{eq-geninv} gives \begin{equation}
\mathcal{D}_sx_\nu=-(A-\lambda_\nu B)^-(\mathcal{D}_sA-\lambda_\nu\mathcal{D}_sB)x_\nu-\tfrac12 (x_\nu'(\mathcal{D}_sB)x_\nu)x_\nu.\label{eq-singvec}
\end{equation} The equations \eqref{eq-singval} and \eqref{eq-singvec}
will be used frequently throughout this paper.

For eigenvalues \eqref{eq-singval} shows that computing
\(\mathcal{D}_s\lambda_\nu\) only requires us to know \(\lambda_\nu\)
and \(x_\nu\), not the other eigenvalues and eigenvectors. Equation
\eqref{eq-singvec} suggests the same thing is true for
\(\mathcal{D}_sx_\nu\), but this is only apparent, because we need all
generalized eigenvalues and eigenvectors to compute
\((A-\lambda_\nu B)^-\). If we only need the derivatives of, say, the
first few eigenvectors from a very large system this can be
computationally quite expensive.

There is an alternative way of computing \(\mathcal{D}_sx_\nu\) which
does not require a full GEVD. We know that \(\mathcal{D}_sx_\nu\) is a
solution of the linear equations \begin{equation}
(A-\lambda_\nu B)\mathcal{D}_sx_\nu=-(\mathcal{D}_sA-\lambda_\nu\mathcal{D}_sB)x_\nu+(\mathcal{D}_s\lambda_\nu)Bx_v. \label{eq-withy}
\end{equation} Because \(\lambda_\nu\) is a simple eigenvalue, the
matrix \(A-\lambda_\nu B\) is of rank \(n-1\), with a null space
consisting of all vectors proportional to \(x_\nu\). Write \(y_\nu\) for
the right hand side of \eqref{eq-withy}. The system is solvable because
\(x_\nu'y_\nu=0\) by \eqref{eq-singval}, and its general solution is
\begin{equation}
\mathcal{D}_sx_\nu=(A-\lambda_\nu B)^-y_\nu+\theta x_\nu,\label{eq-withtheta}
\end{equation} with \((A-\lambda_\nu B)^-\) any generalized inverse of
\(A-\lambda_\nu B\) and with \(\theta\) arbitrary.

Since we must have \begin{equation}
x_\nu'B(\mathcal{D}_sx_\nu)=-\tfrac12 x_\nu'(\mathcal{D}_sB)x_\nu\label{eq-norme}
\end{equation} from \eqref{eq-fromtwo}, we see that \begin{equation}
\theta=-\tfrac12 x_\nu'(\mathcal{D}_sB)x_\nu-x_\nu'B(A-\lambda_\nu B)^-y_\nu.\label{eq-thesolve}
\end{equation} The value of \(\theta\) depends on the choice of the
generalized inverse. For our previous choice of the reflexive inverse
from \eqref{eq-geninv} we actually have
\(x_\nu'B(A-\lambda_\nu B)^-=x_\nu'X^{-T}X^{-1}X(\Lambda-\lambda_\nu I)^+X'y_\nu=e_\nu'(\Lambda-\lambda_\nu I)^+X'y=0\),
and thus \(\theta=-\tfrac12 x_\nu'(\mathcal{D}_sB)x_\nu\), but for other
generalized inverses this may not be true. One possible other choice is
the Moore-Penrose inverse, which we can compute with \begin{equation}
(A-\lambda_\nu B)^+=(A-\lambda_\nu B+\tilde x_\nu \tilde x_\nu')^{-1}-\tilde x_\nu \tilde x_\nu',\label{eq-mpdef}
\end{equation} with \(\tilde x_\nu\) defined as \(x_\nu\) normalized to
length one. Using \eqref{eq-mpdef} means that to compute
\(\mathcal{D}_sx_\nu\) we do not need the complete GEVD of \((A,B)\) but
instead the inverse of the symmetric matrix
\(A-\lambda_\nu B+\tilde x_\nu \tilde x_\nu'\). Yet another generalized
inverse uses one of the non-singular principal submatrices
\(A-\lambda_\nu B\) of order \(n-1\). If \(x_{i\nu}\neq 0\) then the
submatrix leaving out row and column \(i\) is non-singular, so it makes
sense to choose \(i\) with the largest \(|x_{i\nu}|\). The computation
now requires us to invert a matrix of order \(n-1\). Using this last
choice of generalized inverse is known in the engineering literature as
Nelson's method for computing derivatives of eigenvectors, after Nelson
(\citeproc{ref-nelson_76}{1976}).

\subsection{Second Partials}\label{second-partials}

To find second partials of the eigenvalues we differentiate
\eqref{eq-singval} with respect to \(\theta_t\). This gives
\begin{equation}
\mathcal{D}_{st}\lambda_\nu =2x_\nu '(\mathcal{D}_sA-\lambda_\nu \mathcal{D}_sB)\mathcal{D}_tx_\nu +
x_\nu '(D_{st}A-\lambda_\nu\mathcal{D}_{st}B)x_\nu -x_\nu '(\mathcal{D}_sB)x_\nu \cdot\mathcal{D}_t\lambda_\nu .\label{eq-seclbd}
\end{equation} Substituting from \eqref{eq-singval} and
\eqref{eq-singvec} gives, using the reflexive inverse from
\eqref{eq-geninv}, \begin{multline}
\mathcal{D}_{st}\lambda_\nu =-2x_\nu '(\mathcal{D}_sA-\lambda_\nu \mathcal{D}_sB)(A-\lambda_\nu B)^- (\mathcal{D}_tA-\lambda_\nu \mathcal{D}_tB)x_\nu +
x_\nu '(D_{st}A-\lambda_\nu \mathcal{D}_{st}B)x_\nu \\-x_\nu '(\mathcal{D}_tB)x_\nu \cdot x_\nu '(\mathcal{D}_sA-\lambda_\nu \mathcal{D}_sB)x_\nu -x_\nu '(\mathcal{D_s}B)x_\nu \cdot x_\nu '(\mathcal{D}_tA-\lambda_\nu \mathcal{D}_tB)x_\nu .\label{eq-seclbdmat}
\end{multline} Formula \eqref{eq-seclbdmat} shows that
\(\mathcal{D}_{st}\lambda_\nu=\mathcal{D}_{ts}\lambda_\nu\), just as it
should be.

The second partials of the generalized eigenvalues are, not
surprisingly, more complicated. We start with \eqref{eq-secder1} and
differentiate with respect to \(\theta_t\). We give some intermediate
calculations for this case, because they are also used in our software.
First \begin{multline}
\mathcal{D}_{st}x_\nu =-\sum_{\substack{\eta=1\\\eta\neq\nu}}^n\mathcal{D}_t\left\{\frac{x_j'\{\mathcal{D}_sA-\lambda_\nu \mathcal{D}_sB\}x_\nu }{\lambda_\eta-\lambda_\nu }\right\}x_\eta\\
-\sum_{\substack{\eta=1\\\eta\neq\nu}}^n\frac{x_\eta'\{\mathcal{D}_sA-\lambda_\nu \mathcal{D}_sB\}x_\nu }{\lambda_\eta-\lambda_\nu }\mathcal{D}_tx_\eta
-\tfrac12\mathcal{D}_t\{x_\nu '(\mathcal{D}_sB)x_\nu \cdot x_\nu \}.\label{eq-inter1}
\end{multline} The terms in the first summation in \eqref{eq-inter1} are
\begin{multline}
\mathcal{D}_t\left\{\frac{x_\eta'\{\mathcal{D}_sA-\lambda_\nu \mathcal{D}_sB\}x_\nu }{\lambda_\eta-\lambda_\nu }\right\}=\\
\frac{\mathcal{D}_t\{x_\eta'(\mathcal{D}_sA-\lambda_\nu \mathcal{D}_sB)x_\nu \}}{\lambda_\nu -\lambda_\eta}
-\frac{(\mathcal{D}_t\lambda_\eta-\mathcal{D}_t\lambda_\nu )x_\eta'(\mathcal{D}_sA-\lambda_\nu \mathcal{D}_sB)x_\nu }{(\lambda_\eta-\lambda_\nu )^2},\label{eq-inter2}
\end{multline} and the numerator in the first term on the right-hand
side of \eqref{eq-inter2} is \begin{multline}
\mathcal{D}_t\{x_\eta'(\mathcal{D}_sA-\lambda_\nu \mathcal{D}_sB)x_\nu \}=(\mathcal{D}_tx_\eta)'(\mathcal{D}_sA-\lambda_\nu \mathcal{D}_sB)x_\nu +x_\eta'(\mathcal{D}_sA-\lambda_\nu \mathcal{D}_sB)\mathcal{D}_tx_\nu \\+
x_\eta'(\mathcal{D}_{st}A-\lambda_\nu \mathcal{D}_{st}B)x_\nu -\mathcal{D}_t\lambda_\nu \cdot x_\eta'(\mathcal{D}_sB)x_\nu .
\label{eq-inter3}
\end{multline} Finally \begin{multline}
\mathcal{D}_t\{x_\nu '(\mathcal{D}_sB)x_\nu \cdot x_\nu \}=\\2(\mathcal{D}_tx_\nu )'(\mathcal{D}_sB)x_\nu \cdot x_\nu +x_\nu '(\mathcal{D}_{st}B)x_\nu \cdot x_\nu +x_\nu '(\mathcal{D}_tB)x_\nu \cdot\mathcal{D}_tx_\nu .\label{eq-inter4}
\end{multline} In computing the final result we first evaluate
\eqref{eq-inter3}, substitute the result in \eqref{eq-inter2}, and then
substitute that result, together with the result of \eqref{eq-inter4},
into \eqref{eq-inter1}. For this to work we need to have both
\(\mathcal{D}\Lambda\) and \(\mathcal{D}X\) available. We could of
course substitute the expressions of \(\mathcal{D}\Lambda\) and
\(\mathcal{D}X\) into \eqref{eq-inter1}-\eqref{eq-inter4} but that would
result in very long and opaque formulae. We will rely on our software to
give us the numerical values for any specific value of \(\theta\).

\subsection{Multiple Eigenvalues}\label{multiple-eigenvalues}

We would be remiss if we did not say anything about matrices with
multiple eigenvalues, where differentiability fails. The main problem is
that there is no unique eigenvector corresponding with a multiple
eigenvalue. If \(X\) are is an \(n\times r\) matrix of eigenvectors
corresponding with eigenvalue \(\lambda\), i.e.~if \(AX=\lambda BX\),
then \(XS\) are also eigenvectors corresponding with \(\lambda\) for any
\(r\times p\) matrix \(S\).

A related problem has to do with the order of the eigenvalues. Let's
illustrate both problems with a embarrassingly simple example. Suppose
\(B(\theta)=I\) and \[
A(\theta)=\begin{bmatrix}
4&0&0&0\\
0&4&0&0\\
0&0&2&0\\
0&0&0&2
\end{bmatrix}
+\theta
\begin{bmatrix}
4&0&0&0\\
0&3&0&0\\
0&0&2&0\\
0&0&0&1
\end{bmatrix}.
\] The eigenvectors are the four unit vectors
\(e_1-e_4\)\footnote{A unit vector $e_i$ has zeroes everywhere, except for element $i$, which is one.}
The corresponding eigenvalues are \(\lambda_1=4+4\theta\),
\(\lambda_2=4+3\theta\), \(\lambda_3=2+2\theta\), and
\(\lambda_4=2+\theta\). In the figure below they are drawn,
respectively, in RED, BLUE, GREEN, and PURPLE.

\begin{figure}[H]

{\centering \pandocbounded{\includegraphics[keepaspectratio]{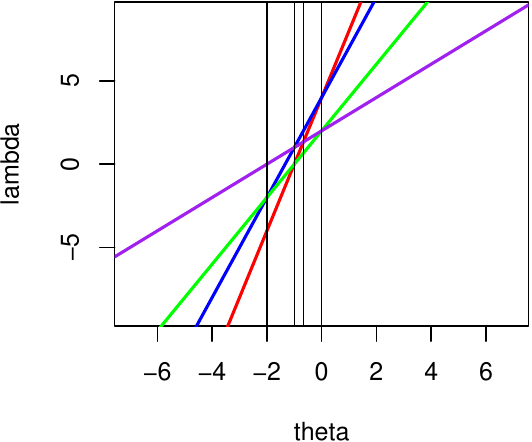}}

}

\caption{Eigenvalues}

\end{figure}%

With this ordering the eigenvalues are clearly differentiable, even
analytic. The corresponding eigenvectors are constant unit vectors,
which also makes them differentiable. If \(A()\) and \(B()\) are
analytic functions of a single parameter \(\theta\) then there are \(n\)
analytic functions \(\lambda_\nu()\) such that \(\lambda_\nu(\theta)\)
is a generalized eigenvalue of the pair \(A(\theta),B(\theta)\) for each
\(\theta\). Moreover there are corresponding analytic \(B\)-orthogonal
eigenvectors \(x_\nu()\) (Sun (\citeproc{ref-sun_90}{1990})).

But if we look at the ordered eigenvalues
\(\lambda_{(1)}\geq...\geq\lambda_{(4)}\) the figure tells us something
different. The largest eigenvalue \(\lambda_{(1)}\) is RED for
\(\theta>0\), BLUE for \(-1<\theta<0\), and PURPLE for \(\theta<-1\). It
is convex, but not differentiable at \(0\) and \(-1\), although it has
left and right derivatives (i.e.~directional derivatives) at all
\(\theta\). The smallest eigenvalue \(\lambda_{(4)}\) is concave, PURPLE
for \(\theta>0\), GREEN for \(-1<\theta<0\), and RED for \(\theta<-1\).
The intermediate eigenvalues \(\lambda_{(2)}\) and \(\lambda_{(3)}\)
change color, and thus slope, even more frequently. At zero none of the
four ordered eigenvalues is differentiable. The situation with
eigenvectors is even worse. At the points where the ordered eigenvalues
change color the eigenvector \(\lambda_{(i)}\) changes discontinuously
from one unit vector to another.

One way to deal with this problem is to give up the idea of looking at
individual eigenvalues and eigenvectors. The averages of blocks of
eigenvalues are differentiable, even if all eigenvalues in the block are
the same (and the ones outside the block are different) (Chu
(\citeproc{ref-chu_90}{1990})). In our example we compute the averages
of the first two and the last two eigenvalues and we see they are
differentiable at zero. The figure below shows the two averages. If
\(\theta=-1\) the average of the first two largest eigenvalues is equal
to the average of the last two, and there are problems again with
differentiability.

\begin{figure}[H]

{\centering \pandocbounded{\includegraphics[keepaspectratio]{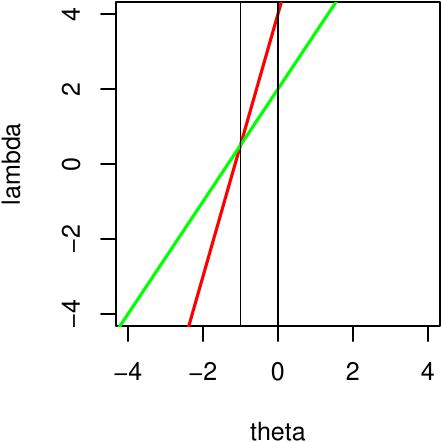}}

}

\caption{Eigenvalue Averages}

\end{figure}%

Unfortunately the result on the analyticity of suitable ordered
eigenvalues and eigenvectors is no longer true if there is more than one
parameter. Consider \[
A(\theta)=\begin{bmatrix}
1+\theta_1&\theta_2\\
\theta_2&1-\theta_1
\end{bmatrix}
\] and \(B(\theta)=I\). The generalized eigenvalues are
\(1\pm\sqrt{\theta_1^2+\theta_2^2}\). The functions
\(\lambda_1(\theta):=1+\|\theta\|_2\) and
\(\lambda_2(\theta):=1-\|\theta\|_2\) are two parabolic shells, the
first one convex, the second one concave. The first one is larger than
the second one for all \(\theta\), except at the origin where the two
shells touch. Thus there is no problem with the ``crossing'' of the two
functions. But neither of the two functions is differentiable at the
origin, although by convexity/concavity they both are directionally
differentiable there.

This points to another solution of part of the problem with multiple
eigenvalues. We can use more general derivatives, such as one-sided
directional derivatives, subdifferentials, epi-derivatives, or
lexicographic derivatives. This is the approach taken in convex and
nonsmooth analysis (Hiriart-Urruty and Ye
(\citeproc{ref-hiriart-urruty_ye_95}{1995}), Lewis and Overton
(\citeproc{ref-lewis_overton_96}{1996})). It has been applied mainly to
eigenvalues in EVD problems with \(B=I\), and not to the corresponding
eigenvectors. Eigenvalues have a variational characterization and
eigenvectors do not. The basic result from this approach is that
eigenvalues, either simple or multiple, are directionally differentiable
functions of the matrix. There are different definitions around of
second-order differentiability, but for the most common ones eigenvalues
are twice directionally differentiable (Torki
(\citeproc{ref-torki_99}{1999}), Torki (\citeproc{ref-torki_01}{2001})).

\sectionbreak

\section{Software}\label{software}

In the
\href{https://www.github.com/deleeuw/generalized-eigen-derivatives}{github
repository} there are R programs evaluating
\(A, B, \mathcal{D}\Lambda, \mathcal{D}X, \mathcal{D}^2\Lambda,\) and
\(\mathcal{D}^2X\) at a given \(\theta\). The function gevdNonlinear()
can be used for any non-linear GEVD. Its only argument is the vector
\(\theta\). It computes the first partials and then calls the
subroutines gevdHessianValues() and gevdHessianVectors() to compute the
second partials. The gevdHessianVectors() routine uses the stepwise
method suggested at the end of the previous section. First compute all
first derivatives, and then use their numerical values in equations
\eqref{eq-inter1}-\eqref{eq-inter4}.

In order to attain the necessary level of generality gevNonlinear()
needs to be run from or after a template which defines global variables,
specifically the R functions theA(), theB(), dA(), dB(), ddA(), and
ddB(), together with the parameters and additional values that these
auxiliary functions need. The template also has values for

\begin{itemize}
\tightlist
\item
  \(p\), the number of parameters;
\item
  \(n\), the order of the matrices;
\item
  hessianl, do we compute second derivatives of eigenvalues;
\item
  hessianx, do we compute second derivatives of eigenvectors.
\end{itemize}

If only first order information is needed we can set hessianl and
hessianx to FALSE.

The result of gevdNonlinear() is a list with

\begin{itemize}
\tightlist
\item
  \(A(\theta)\), an \(n\times n\) matrix, typical element
  \(a_{ij}(\theta)\),
\item
  \(B(\theta)\), an \(n\times n\) matrix, typical element
  \(b_{ij}(\theta)\),
\item
  \(\lambda(\theta)\), an \(n\)-element vector, typical element
  \(\lambda_\nu(\theta)\),
\item
  \(X(\theta)\), an \(n\times n\) matrix, typical element
  \(x_{i\nu}(\theta)\),
\item
  \(\mathcal{D}\Lambda(\theta)\), an \(n\times p\) matrix, element
  \((\nu, s)\) is \(\mathcal{D}_s\lambda_\nu(\theta)\),
\item
  \(\mathcal{D}X(\theta)\), an \(n\times p\times n\) array, element
  \((i, s, \nu)\) is \(\mathcal{D}_sx_{i\nu}(\theta)\),
\item
  \(\mathcal{D}^2\Lambda(\theta)\), a \(p\times p\times n\)
  array,element \((s, t, \nu)\) is
  \(\mathcal{D}_{st}\lambda_\nu(\theta)\),
\item
  \(\mathcal{D}^2X(\theta)\), a \(p\times p\times n\times n\) array,
  element \((s, t, i, \nu)\) is \(\mathcal{D}_{st}x_{i\nu}(\theta)\).
\end{itemize}

The names in the code and in the results for the eight matrices in this
list are a, b, l, x, dl, dx, ddl, and ddx. The R print function prints
\(\mathcal{D}\Lambda\) as an \(n\times p\) matrix and \(\mathcal{D}X\)
as a sequence of \(n\) matrices of dimensions \(n\times p\), one for
each eigenvalue. \(\mathcal{D}^2\Lambda\) is printed as \(n\) symmetric
matrices of order \(p\), and \(\mathcal{D}^2X\) as \(\smash{n^2}\)
symmetric matrices of order \(p\), one for each element of \(X\).
Printing \(\mathcal{D}^2X\) produces a lot of output if \(n\) and \(p\)
are at all large. If hessianx and hessianl are set to FALSE then ddl and
ddx are set to NULL.

There is a corresponding numDeriv based check function
gevdNonlinearNum(), which has the same single argument \(\theta\) and
gives the same output as gevdNonlinear(). The same procedure is followed
for evdNonlinear() and gsvdNonlinear(), which are the functions for the
(non-generalized) eigenvalue decomposition and for the generalized
singular value decomposition we shall discuss in later sections.

\sectionbreak

\section{Template Examples}\label{template-examples}

The function gevNonlinear() can handle any parametric model, as long as
the appropriate template is provided with routines for
\(A, B, \mathcal{D}A, \mathcal{D}B, \mathcal{D}^2A,\) and
\(\mathcal{D}^2B\). In special cases more compact and elegant formulae
and more efficient computations will exist. The question then becomes if
it is worthwhile to derive and retype the formulae and to reprogram the
R routines for these special cases. In this section we discuss some
important special cases and indicate what should be included in the
templates.

We do not give the numerical results of the various derivative
calculations, which for these examples are just a large amount of
meaningless numbers. The software in the repository, together with the
templates, can be used to calculate Jacobians and Hessians for various
values of the parameter vector. I should mention that in all examples
the results for the derivatives based on the formulas are the same as
the results based on numerical differentiation.

\subsection{Power Series}\label{sec-single}

In our first example we define \begin{align}
A(\theta)&=A_0+\theta A_1+\frac12\theta^2A_2+...+\frac{1}{p}\theta^pA_p\\
B(\theta)&=B_0+\theta B_1+\frac12\theta^2B_2+...+\frac{1}{p}\theta^pB_p
\end{align} By a suitable choice of the \(A_s\) and \(B_s\) we can
approximate functions for the cells of \(A()\) and \(B()\) with
convergent power series expansions (possibly different functions for
different cells).

The template for a case with \(n=4\) and \(p=3\) is
gevdTemplatePowerSeries.R in the github repository. U=It uses random
symmetric matrices for \(A_s\) and \(B_s\).

\subsection{Linear Combinations}\label{sec-linear}

Suppose \(A()\) is a linear combination of \(p\) known symmetric
matrices \(A_s\) and \(B()\) is a linear combination of \(p\) known
symmetric matrices \(B_s\). There may also be ``intercepts'' \(A_0\) and
\(B_0\). Thus \begin{subequations}
\begin{align}
A(\theta)&=A_0+\sum_{s=1}^p\theta_sA_s,\\
B(\theta)&=B_0+\sum_{s=1}^p\theta_sB_s.
\end{align}
\end{subequations} It may seem somewhat limiting that the same parameter
vector \(\theta\) is used for both \(A()\) and \(B()\). But we can
decide to make the last \(q\) matrices \(A_s\) and the first \(p-q\)
matrices \(B_s\) equal to zero. This means, in effect, that \(A()\) and
\(B()\) depend on different parameter vectors.

The template for a special case with \(n=4\) and \(p=6\) is
gevdTemplateLinear.R in the
\href{https://www.github.com/deleeuw/generalized-eigen-derivatives}{github
repository}. In this example the last three \(A_s\) and the first three
\(B_s\) are indeed zero. The basic simplication is of course that
\(\mathcal{D}_sA=A_s\) and \(\mathcal{D}_sB=B_s\), and that consequently
the second partials \(\mathcal{D}_{st}A\) and \(\mathcal{D}_{st}B\) are
all zero.

\subsection{The Elements}\label{the-elements}

As a special case of thus linear combinations special case we have the
partial derivatives with respect to the elements of \(A\) and \(B\).
Differentiation with respect to the matrix elements can be useful in
combination with the chain rule, which computes, for example, the
derivative of the eigenvalue with respect to a parameter as the
derivative of the eigenvalue with respect to the matrix evaluated at the
derivative of the matrix with respect to the parameter.

The linear combinations we use in this case are \begin{subequations}
\begin{align}
A&=\mathop{\sum\sum}_{1\leq i\leq j}a_{ij}E_{ij},\\
B&=\mathop{\sum\sum}_{1\leq i\leq j}b_{ij}E_{ij}.
\end{align}
\end{subequations} Here \(E_{ij}:=(e_ie_j'+e_je_i')\) for \(i\neq j\)
and \(E_{ii}=e_ie_i'\) with \(e_i\) and \(e_j\) unit vectors. Thus
\begin{equation}
\mathcal{D}_{(i,j)}A=\mathcal{D}_{(i,j)}B=E_{ij}.
\end{equation}

In gevdTemplateLinear.R from the previous secton the matrices \(A_s\)
and \(B_s\) were arbitrary and were explicitly given as input. In
gevdTemplateElement.R there are auxiliary functions that generate the
matrices \(E_{ij}\) when needed. This requires some manipulationof
indices to go from the vector of parameters \(\theta_s\) to the matrices
of parameters \(a_{ij}\) or \(b_{ij}\). The included numerical example
is of order four, and consequently has 20 parameters.

\subsection{Eigenvalue Decomposition}\label{eigenvalue-decomposition}

An Eigenvalue Decomposition (EVD) problem, without the ``generalized'',
is of the form \begin{subequations}
\begin{align}
AX&=X\Lambda,\\
X'X&=I.
\end{align} 
\end{subequations} This is the special case of GEVD with \(B=I\) and
\(\mathcal{D}B=0\). In EVD the matrices \((A-\lambda_\nu B)^-\) from
\eqref{eq-geninv} are actually the Moore-Penrose inverses of
\(A-\lambda_\nu I\). There are substantial, and rather obvious,
simplifications of the first and second derivatives.

From \eqref{eq-singval} \begin{equation}
\mathcal{D}_s\lambda_\nu=x_\nu'(\mathcal{D}_sA)x_\nu,\label{eq-sevalsim}
\end{equation} and from \eqref{eq-singvec} \begin{equation}
\mathcal{D}_sx_\nu=-(A-\lambda_\nu I)^+(\mathcal{D}_sA)x_\nu.\label{eq-vecpersimmp}
\end{equation}

From \eqref{eq-seclbdmat} \begin{equation}
\mathcal{D}_{st}\lambda_\nu=-2x_\nu'(\mathcal{D}_sA)(A-\lambda_\nu I)^+(\mathcal{D}_tA)x_\nu+
x_\nu '(D_{st}A)x_\nu .\label{eq-hessvalsimple}
\end{equation} An alternative, which may be more efficient
computationally, is \[
\mathcal{D}_{st}\lambda_\nu=-2\sum_{\substack{\eta=1\\\eta\neq\nu}}^n\frac{1}{\lambda_\eta-\lambda_\nu}x_\nu'(\mathcal{D}_sA)x_\eta\cdot x_\nu'(\mathcal{D}_tA)x_\eta+x_\nu'(\mathcal{D}_{st}A)x_\nu.
\] This is also the formula given in Overton and Womersley
(\citeproc{ref-overton_womersley_95}{1995}).

For the second derivatives of the eigenvectors we again use the stepwise
computation of \eqref{eq-inter1}. We do not give a single formula, but a
recipe instead. \begin{equation}
\mathcal{D}_{st}x_\nu =-\sum_{\substack{\eta=1\\\eta\neq\nu}}^n\mathcal{D}_t\left\{\frac{x_\eta'\{\mathcal{D}_sA\}x_\nu }{\lambda_\eta-\lambda_\nu }\right\}x_\eta
-\sum_{\substack{\eta=1\\\eta\neq\nu}}^n\frac{x_\eta'\{\mathcal{D}_sA\}x_\nu }{\lambda_\eta-\lambda_\nu }\mathcal{D}_tx_\eta.\label{eq-jnter1}
\end{equation} The terms in the first summation in \eqref{eq-jnter1} are
\begin{equation}
\mathcal{D}_t\left\{\frac{x_\eta'\{\mathcal{D}_sA\}x_\nu }{\lambda_\eta-\lambda_\nu }\right\}=
\frac{\mathcal{D}_t\{x_\eta'(\mathcal{D}_sA)x_\nu \}}{\lambda_\nu -\lambda_\eta}
-\frac{(\mathcal{D}_t\lambda_\eta-\mathcal{D}_t\lambda_\nu )x_\eta'(\mathcal{D}_sA)x_\nu }{(\lambda_\eta-\lambda_\nu )^2},\label{eq-jnter2}
\end{equation} and the numerator in the first term of \eqref{eq-jnter2}
is \begin{equation}
\mathcal{D}_t\{x_\eta'(\mathcal{D}_sA)x_\nu \}=(\mathcal{D}_tx_\eta)'(\mathcal{D}_sA)x_\nu +x_\eta'(\mathcal{D}_sA)\mathcal{D}_tx_\nu
+x_\eta'(\mathcal{D}_{st}A)x_\nu.\label{eq-jnter3}
\end{equation} As before, we substitute \eqref{eq-jnter3} into
\eqref{eq-jnter2} and then substitute the result in \eqref{eq-jnter1}.

It is easy and not very wasteful to plug in \(B=I\) and
\(\mathcal{D}_sB=\mathcal{D}_{st}B=0\) in the formulas and the template
of the generalized nonlinear case. Nevertheless, because of the
importance of EVD, we do include both evdTemplate.R and evdNonlinear.R
in the repository. The template has a one-parameter example
\(A(\theta)=A_0+\theta A_1\) of order four.

\subsection{Generalized Singular Value
Decomposition}\label{generalized-singular-value-decomposition}

Suppose \(F\) is a rectangular \(n\times m\) matrix, and \(G\) and \(H\)
are two positive definite matrices of orders \(n\) and \(m\). Without
loss of generality we assume \(m\leq n\). The Generalized Singular Value
Decomposition (GSVD) is finding a solution to the system
\begin{subequations}
\begin{align}
FY&=GX\Lambda,\label{eq-gsvdef1a}\\
F'X&=HY\Lambda,\label{eq-gsvdef2a}\\
X'GX&=I,\label{eq-gsvdef3a}\\
Y'HY&=I,\label{eq-gsvdef4a}
\end{align}
\end{subequations} Here \(Y\) and \(\Lambda\) are square of order \(m\),
with \(\Lambda\) diagonal and non-negative. The left singular vectors
are \(n\times m\), and there is an \(n\times(n-m)\) matrix \(X_\perp\)
that satisfies \(F'X_\perp=0\), with \(X_\perp'GX=0\) and
\(X_\perp'GX_\perp=I\).

The GSVD system \eqref{eq-gsvdef1a}-\eqref{eq-gsvdef4a} is (very)
closely related to the GEVD system \begin{equation}
\begin{bmatrix}
0&F\\F'&0
\end{bmatrix}
\begin{bmatrix}
U\\V
\end{bmatrix}
=
\begin{bmatrix}
G&0\\
0&H
\end{bmatrix}
\begin{bmatrix}
U\\V
\end{bmatrix}
\Psi,\label{eq-gsvdgevd}
\end{equation} with normalization \(U'GU+V'HV=I\).

System \eqref{eq-gsvdgevd} has \(n+m\) solutions that can be described
using the GSVD solutions \((X,Y,\Lambda)\). There are \(m\) solutions
with \(U=\tfrac12\sqrt{2}\ X, V=\tfrac12\sqrt{2}\ Y\) and
\(\Psi=\Lambda\). There are another \(m\) solutions with
\(U=\tfrac12\sqrt{2}\ X, V=-\tfrac12\sqrt{2}\ Y\) and \(\Psi=-\Lambda\).
And finally there are \(n-m\) solutions with \(U=X_\perp, V=0\), and
\(\Lambda=0\). For the non-zero eigenvalues we have
\(u_\nu'Gu_\nu=v_\nu'Hv_\nu=\tfrac12\), which shows that the GSVD
solutions are normalized slightly different from the GEVD ones. The
\(m\) solutions with non-negative eigenvalues are the interesting ones,
and they provide us with the \(m\) solutions of the GSVD.

Thus, in stead of tackling the GSVD from
\eqref{eq-gsvdef1a}-\eqref{eq-gsvdef4a} directly, we apply our results
for the derivatives of the GEVD to the system \eqref{eq-gsvdgevd}. All
three matrices \(F,G,\) and \(H\) are assumed to be functions of the
parameters \(\theta\) and we will only consider the first \(m\)
generalized eigenvalues (assuming that \(F\) has rank \(r=m\), otherwise
only the first \(r\)).

To bring the notation in line with our previous results we rewrite the
equations in the GEVD system \eqref{eq-gsvdgevd} as \begin{equation}
A\begin{bmatrix}x_\nu\\y_\nu\end{bmatrix}=\lambda_\nu B\begin{bmatrix}x_\nu\\y_\nu\end{bmatrix},
\end{equation} with \begin{equation}
A=\begin{bmatrix}
0&F\\F'&0
\end{bmatrix}
\end{equation} and \begin{equation}
B=\begin{bmatrix}
G&0\\
0&H
\end{bmatrix}
\label{eq-gsvdgevd2}
\end{equation}

The first derivatives are \begin{equation}
\mathcal{D}_s\lambda_\nu=2\ x_\nu'(\mathcal{D}_sF)y_\nu-\lambda_\nu (x_\nu'(\mathcal{D}_sG)x_\nu+y_\nu'(\mathcal{D}_sH)y_\nu),\label{eq-gsvdderlab}
\end{equation} and \begin{multline}
\mathcal{D}_s\begin{bmatrix}x_\nu\\y_\nu\end{bmatrix}=-(A-\lambda_\nu B)
^-
\begin{bmatrix}
(\mathcal{D}_sF)y_\nu-\lambda_\nu(\mathcal{D}_sG)x_\nu\\
(\mathcal{D}_sF')x_\nu-\lambda_\nu(\mathcal{D}_sH)y_\nu\end{bmatrix}\\
-\tfrac12 
(x_\nu'(\mathcal{D}_sG)x_\nu+y_\nu'(\mathcal{D}_sH)y_\nu)\cdot\begin{bmatrix}x_\nu\\y_\nu\end{bmatrix}.
\label{eq-gsvdder1}
\end{multline} Note that \begin{equation}
(A-\lambda_\nu B)^-=\sum_{\substack{\eta=1\\\eta\neq\nu}}^{n+m}\frac{1}{\lambda_\eta-\lambda_\nu}\begin{bmatrix}x_\eta x_\eta'&x_\eta y_\eta'\\y_\eta x_\eta'&y_\eta y_\eta'\end{bmatrix}\label{eq-expgen}
\end{equation} with the summation over all \(n+m\) eigenvalues,
including \(m\) negative ones and the \(n-m\) that are equal to zero.

It is clearly helpful to define \begin{subequations}
\begin{align}
T_s&:=X'(\mathcal{D}_sF)Y,\label{eq-helpa}\\
U_s&:=X'(\mathcal{D}_sG)X,\label{eq-helpb}\\
V_s&:=Y'(\mathcal{D}_sH)Y.\label{eq-helpc}
\end{align}
\end{subequations} With definitions \eqref{eq-helpa}-\eqref{eq-helpc}
the eigenvalue derivatives \eqref{eq-gsvdderlab} become \begin{equation}
\mathcal{D}_s\lambda_\nu=2\ t^s_{\nu\nu}-\lambda_\nu (u^s_{\nu\nu}+v^s_{\nu\nu}).
\end{equation} Substitute \eqref{eq-expgen} into \eqref{eq-gsvdder1},
and use the definitions \eqref{eq-helpa}-\eqref{eq-helpc} for the
eigenvector derivatives. \begin{align}
\mathcal{D}_sx_\nu&=-\sum_{\substack{\eta=1\\\eta\neq\nu}}^{n+m}\frac{1}{\lambda_\eta-\lambda_\nu}(t^s_{\eta\nu}+t^s_{\nu\eta}-\lambda_\nu(u^s_{\eta\nu}+v^s_{\nu\eta}))x_\eta-\tfrac12 
(u^s_{\nu\nu}+v^s_{\nu\nu})x_\nu\\
\mathcal{D}_sy_\nu&=-\sum_{\substack{\eta=1\\\eta\neq\nu}}^{n+m}\frac{1}{\lambda_\eta-\lambda_\nu}(t^s_{\eta\nu}+t^s_{\nu\eta}-\lambda_\nu(u^s_{\eta\nu}+v^s_{\nu\eta}))y_\eta-\tfrac12 
(u^s_{\nu\nu}+v^s_{\nu\nu})y_\nu
\end{align}

In a Singular Value Decomposition (SVD) we have both \(G=I\) and
\(H=I\), and the corresponding derivatives are consequently zero, as are
the matrices \(U_s\) and \(V_s\). This simplifies matters considerably.

The repository has the template gsvdTemplate.R, which can be used
together with gevdNonlinear() to compute first and second derivatives of
singular values and vectors. The example in the template is a simple one
with six parameters. \(F(\theta)\) is \(4\times 3\), with the columns
\((\theta, \theta^2, \theta^3)\). That requires four parameters. The
remaining two parameters are used for \(G\) and \(H\), which are both of
the form \(I+\theta(I-ee')\). Thus they have one on the diagonal and
\(\theta\) on the off-diagonal. The additional file gsvdNonlinear.R has
the function gsvdNonlinear(), with its numDeriv checker, which computes
first derivatives of singular values and vectors. It uses the formulas
in this section and is less computationally wasteful as gevdNonlinear().
For now, it does not compute second derivatives,

\subsection{Correspondence Analysis}\label{correspondence-analysis}

The next two examples (correspondence analysis and multiple
correpondence analysis) can be used in combination with the Delta
Method. Again, the Delta Method literature is huge and diverse. Special
cases have been around before there was a statistics discipline, ever
since the beginning of error analysis in geodesy, physics, and astronomy
(Gorroochurn (\citeproc{ref-gorroochurn_20}{2020})). For the types of
applications we have in mind all the relevant Delta Method details are
given in Hsu (\citeproc{ref-hsu_49}{1949}) and Hurt
(\citeproc{ref-hurt_76}{1976}).

Suppose \(A()\) and \(B()\) are functions of proportions based on \(n\)
iid observations. We use the Dutch Convention of underlining (sequences
of) random variables (Hemelrijk (\citeproc{ref-hemelrijk_66}{1966})).
The sample proportions are \(\smash{{\underline{p}}_n}\), and their
expected values are \(\pi\). We have asymptotic normality
\begin{equation}
n^{\tfrac12}(\underline{p}_n-\pi)\ \mathop{\rightarrow}^L\ \mathcal{N}(0,\ \Pi-\pi\pi'),
\end{equation} where \(\Pi\) is a diagonal matrix with \(\pi\) on the
diagonal. If \(f\) is a three times continuously differentiable function
with values in \(\mathbb{R}^m\) then \begin{equation}
n^{\tfrac12}(\underline{p}_n-\pi)\ \mathop{\rightarrow}^L\ \mathcal{N}(0,\ \mathcal{D}f(\pi)'(\Pi-\pi\pi')\mathcal{D}f(\pi)).\label{eq-deltavar}
\end{equation} Moreover for the expected value we have the convergence
in probability \begin{equation}
n\mathbf{E}(f_\nu(\underline{p}_n)-f_\nu(\pi))\ \mathop{\rightarrow}^P
\tfrac12\text{tr}\ \mathcal{D}_\nu^2f(\pi)(\Pi-\pi\pi').\label{eq-deltabias}
\end{equation} Plugin versions of formulae \eqref{eq-deltavar} and
\eqref{eq-deltabias}, with the \(\pi\) on the right hand side replace by
\(\smash{\underline{p}_n}\), allow us to use the first and second
derivatives of the eigenvalues and eigenvectors to compute asymptotic
estimates of biases, standard errors, and confidence intervals.

Correspondence analysis is a GSVD, with \(F\) the contingency table,
\(G\) a diagonal matrix with row sums, and \(H\) a diagonal matrix with
column sums. All three matrices are linear functions of the proportions.
Thus \(\mathcal{D}F\) is a unit
matrix\footnote{A unit matrix has one element equal to one, the others are zero.},
and \(\mathcal{D}G\) and \(\mathcal{D}H\) are diagonal unit matrices
with row and columns sums of \(\mathcal{D}F\). All second derivatives
are zero.

The file gevdTemplateCA.R uses the GEVD results for the linear case and
applies them to a social mobility contingency table from Glass
(\citeproc{ref-glass_54}{1954}), also used in De Leeuw and Mair
(\citeproc{ref-deleeuw_mair_A_09b}{2009}). In this 7 × 7 table the
occupational status of fathers (rows) and sons (columns) of 3497 British
families were cross-classiﬁed. The categories are professional and high
administrative, managerial and executive, higher supervisory, lower
supervisory, skilled manual and routine non-manual, semi-skilled manual,
and unskilled manual. Running the template followed by either
gevdNonlinear() or gsvdNonlinear() will give the derivatives needed for
Delta Method standard errors and confidence intervals.

\subsection{Multiple Correspondence
Analysis}\label{multiple-correspondence-analysis}

In Multiple Correspondence Analysis (MCA) we have \(N\) observations on
\(m\) categorical variables (also known as ``factors''). Variable \(j\)
has \(k_j\) possible values (also known as ``levels''), and our
observations are coded as unit vectors of length \(k_j\), concatenated
to binary vectors (also known as ``profiles'') of length
\(K:=\sum k_j\). There are \(M:=\prod k_j\) possible observations, and
each of them occurs in the data with relative frequency \(p_s\) (in most
MCA applications \(N<<M\) and thus many \(\pi_s\) will be zero). Note
that (ordinary) Correspondence Analysis (CA) is the special case with
\(m=2\).

The matrices \(A\) and \(B\) are of the form \begin{subequations}
\begin{align}
A(p)&=\sum_{s=1}^M p_s g_s g_s',\\
B(p)&=m\sum_{s=1}^M p_s\ \text{diag}(g_s g_s').
\end{align}
\end{subequations} The template gevdTemplateMCA.R use the
three-dimensional contingency table HairEyeColor from the R datasets
package, which is a part of base R (R Core Team
(\citeproc{ref-r_core_team_25}{2025})). The first variable is hair color
(4 levels), the second is 2 is eye color (4 levels), and the third
variable is sex (2 levels). There are 592 cross-classified individuals,
students in an introductory statistics course at the University of
Delaware. Thus \(N=592\), \(M=4\times 4\times 2=32\), \(K=4+4+2=10\).

It is abundantly clear that this example should really be analyzed with
some special purpose software which does not require computing and
storing the \(M\) arrays \(A_s\) and the \(M\) arrays \(B_s\), which are
symmetric, integer, and very sparse. In this case we want to show,
however, that gevNonlinear() can do it all. The template
gevdExampleMCA.R can actually handle any three-dimensional table. After
running the template, we apply gevNonlinear().

The results for MCA can also be applied directly to the situation in
which the possible data are a finite number of numerical vectors which
occurs with different relative frequencies. Because of the limits of
measurement precision one could argue that all multivariate data are
actually of this type. But of course for so-called ``continuous'' data
the number of possible values, although finite, can be enormously large.
One solution for the discrete-continuous dilemma is to move from
expectations that are weighted by relative frequencies to expectations
that are weighted by the empirical distribution function. Our formulae
for the derivatives can also be applied to get standard errors,
confidence intervals, and bias corrections for the EVD of covariances
and correlations using the infinitesimal versions of the bootstrap or
the jackknife.

\subsection{Unweighted Least Squares
CCFA}\label{unweighted-least-squares-ccfa}

The final three examples are more complicated. They are all versions of
exploratory common factor analysis (ECFA). In Covariance Exploratory
Common Factor Analysis (CECFA) we approximate a covariance or
correlation matrix by the sum of a diagonal matrix and a matrix of small
rank. There are many different ways in which the degree of approximation
can be measured, and for a given loss function there are many different
proposed algorithms to minimize it. In the next two examples we will
compute the first and second derivatives to provide a basis for
minimization by Newton's method of some of these loss functions.

In Least Squares Factor Analysis (LSFA) of a covariance matrix \(C\) of
order \(n\) we minimize \begin{equation}
\sigma(D)=\sum_{i=p+1}^n\lambda_\nu^2(C-D)\label{eq-ulsloss}
\end{equation} over diagonal matrices \(D\). In a proper LSFA we require
\(D\geq 0\) and \(C-D\) positive semi-definite, but these constraints
need not bother us here because we are only interested in the first and
second derivatives. The formulas we derive have been given earlier by
Derflinger (\citeproc{ref-derflinger_69}{1969}) and Joreskog and Van
Thillo (\citeproc{ref-joreskog_vanthillo_71}{1971}).

Because \(A=C-D\) we have \(\mathcal{D}_i A=-e_ie_i'\). It follows
directly from \eqref{eq-singval} that \begin{equation}
\mathcal{D}_i\sigma=-2\sum_{\nu=p+1}^n\lambda_\nu^{\ }x_{i\nu}^2.\label{eq-duls}
\end{equation} From \eqref{eq-seclbdmat}, taking into account that
\(B=I\), and thus \(D_iB=D_{ij}B=0\), \begin{equation}
\mathcal{D}_{ij}\sigma_\star=2\sum_{\nu=p+1}^nx_{i\nu}^2x_{j\nu}^2-4\sum_{\nu=p+1}^n\sum_{\substack{\eta=1\\\eta\neq\nu}}^n\frac{\lambda_\nu}{\lambda_\eta-\lambda_\nu}x_{i\nu}x_{j\nu}x_{i\eta}x_{j\eta}.\label{eq-dduls}
\end{equation}

The file lsfa.R in the github repository has functions to compute the
first and second derivatives, with the usual numDeriv checker.

\subsection{The Swain CCFA Family}\label{the-swain-ccfa-family}

In an important paper Swain (\citeproc{ref-swain_75}{1975}) derived a
class of factor analytic procedures which produce estimates that are
asymptotically equivalent to the classical Lawley-Joreskog maximum
likelihood estimates. Swain also wrote down the necessary first and
second order derivatives that can be used in implementing Newton's
method, thus generalizing the work of Jennrich and Robinson
(\citeproc{ref-jennrich_robinson_69}{1969}), Derflinger
(\citeproc{ref-derflinger_69}{1969}), and Clarke
(\citeproc{ref-clarke_70}{1970}) for the maximum likelihood method.

Suppose \(\mathcal{F}\) is the set of twice-differentiable univariate
functions defined on the positive real axis with the properties
\begin{subequations}
\begin{align}
f(1)&=\mathcal{D}f(1)=0,\label{eq-f1}\\
\mathcal{D}^2f''(1)&=1,\label{eq-f2}\\
\mathcal{D}f(\theta)&<0\text{ for }0<\theta<1,\label{eq-f3}\\
\mathcal{D}f(\theta)&>0\text{ for }\theta>1.\label{eq-f4}
\end{align}
\end{subequations} Note that this implies that each \(f\in\mathcal{F}\)
is unimodal and has a single minimum equal to zero at one. Also, by
l'Hôpital, \begin{equation}
\lim_{x\rightarrow 1}\frac{f(x)}{\frac12(x-1)^2}=1.
\end{equation} Define the signature of a real-valued function \(f\) at
\(x\) to be -1 if \(f(x)<0\), \(+1\) if \(f(x)>0\), and \(0\) if
\(f(x)=0\). It follows from \eqref{eq-f1}-\eqref{eq-f4} that if
\(f,g\in\mathcal{F}\) then the signatures of \(\mathcal{D}f\) and
\(\mathcal{D}g\) are the same.

Swain proposes to minimize \begin{equation}
\sigma(D):=\sum_{\nu=p+1}^n f(\lambda_\nu(A)),
\end{equation} with \(\lambda\) the eigenvalues of
\(A:=S^{-\frac12}DS^{-\frac12}\). These are also the eigenvalues of the
asymmetric matrix \(S^{-1}D\), and the generalized eigenvalues of the
pair \((D,S)\).

Maximum likelihood estimation, generalized least squares estimation
(Joreskog and Goldberger (\citeproc{ref-joreskog_goldberger_72}{1972})),
and the ``geodesic distance'' between covariance matrices (James
(\citeproc{ref-james_73}{1973})) are special cases. For maximum
likelihood \begin{subequations}
\begin{equation}
f_1(\theta):=1/\theta+\log\theta-1,
\end{equation}
for GLS with loss $\frac12\text{tr}\ S^{-1}(\Sigma-S)S^{-1}(\Sigma-S)$ we have
\begin{equation}
f_2(\theta):=\frac12(\theta-1)^2,
\end{equation}
and for the "geodetic distance" 
\begin{equation}
f_3(\theta):=\frac12(\log\theta)^2.
\end{equation}
But \eqref{eq-f1}-\eqref{eq-f4} also covers other cases
such as the  GLS variant with loss $\frac12\text{tr}\ \Sigma^{-1}(\Sigma-S)\Sigma^{-1}(\Sigma-S)$ which has
\begin{equation}
f_4(\theta):=\frac12\frac{(\theta-1)^2}{\theta^2}.
\end{equation}
\end{subequations} In addition if \(f(\theta)\) satisfies
\eqref{eq-f1}-\eqref{eq-f4} for all \(\theta\) then so does
\(f(\theta^{-1})\). In fact \(f_4(\theta)=f_2(\theta^{-1})\). Also
\(f_3(theta)=f_3(\theta^{-1})\). Note that the unweighted least squares
method from the previous section is not in the Swain family, but the GLS
methods are, and they have the advantage of being scale-free.

From \(A=S^{-\frac12}DS^{-\frac12}\) we have
\(\mathcal{D}_iA=S^{-\frac12}e_ie_i'S^{-\frac12}\). To apply the EVD
derivative formulas it is convenient to define the matrix
\(Y:=S^{-\frac12}X\) with elements \(y_{i\nu}\). It follows that
\begin{subequations}
\begin{equation}
\mathcal{D}_i\lambda_\nu=(x_\nu'S^{-\frac12}e_i)^2=y_{i\nu}^2.
\end{equation}
and
\begin{equation}
\mathcal{D}_ix_\nu=-\sum_{\substack{\eta=1\\\eta\neq\nu}}^n\frac{1}{\lambda_\eta-\lambda_\nu}x_\eta x_\eta'(S^{-\frac12}e_ie_i'S^{-\frac12})x_\nu=-\sum_{\substack{\eta=1\\\eta\neq\nu}}^n\frac{1}{\lambda_\eta-\lambda_\nu} y_{i\nu}y_{i\eta}\cdot x_\eta
\end{equation}
\end{subequations} Consequently \begin{equation}
\mathcal{D}_i\sigma=\sum_{\nu=p+1}^n\mathcal{D}f(\lambda_\nu)\mathcal{D}_i\lambda_\nu=\sum_{\nu=p+1}^n\mathcal{D}f(\lambda_\nu)y_{i\nu}^2.
\end{equation} Differentiate once again \begin{equation}
\mathcal{D}_{ij}\sigma=\sum_{\nu=p+1}^n\mathcal{D}^2f(\lambda_\nu)y_{j\nu}^2y_{i\nu}^2-2\sum_{\nu=p+1}^n\sum_{\substack{\eta=1\\\eta\neq\nu}}^n\mathcal{D}f(\lambda_\nu)\{\frac{1}{\lambda_\eta-\lambda_\nu} y_{i\nu}y_{i\eta}y_{j\nu}y_{j\eta}\}
\end{equation}

The github repository has a file swain.R which has functions to compute
the first and second derivatives for the more common choices of
\(f\in\mathcal{F}\). There is also a numDeriv checker. The functions are
programmed in such a way that they can easily be extended to other loss
functions with \(f\) satisfying \eqref{eq-f1}-\eqref{eq-f4}.

\subsection{Matrix Decomposition Factor
Analysis}\label{matrix-decomposition-factor-analysis}

In Matrix Decomposition Factor Analysis (MDFA) we have to maximize the
trace or nuclear norm \(\|.\|_\tau\), i.e.~the sum of the singular
values, of a matrix \(XT\) over \(T\). This is equivalent to maximizing
the sum of the square roots of the eigenvalues of \(T'X'XT\).

In MDFA the matrix \(X\) is \(n\times m\) and \(T\) is \(m\times p\),
with \(m<p\). Thus \(A=T'X'XT\) is of order \(p\) and of rank
\(m\leq p\). We suppose it is locally of constant rank \(m\) and we
maximize the sum of the square roots of its \(m\) non-zero eigenvalues
(which is equal to the trace norm).

The matrix \(T\) can be patterned, which means that some of its elements
are fixed at known values, usually zero. In exploratory MDFA we
typically have \(T=\begin{bmatrix}U&D\end{bmatrix}\), with an
\(m\times r\) with \(r=p-m\), matrix \(U\) of common factor loadings and
a diagonal matrix \(D\) of uniquenesses. Thus the pattern in this case
is that the off-diagonal elements of \(D\) are zero. In confirmatory
MDFA some elements of \(U\) are fixed at known values as well.

So \(A=T'CT\), with \(C=X'X\) a constant matrix, and
\(T(\theta)=T_0+\sum_{s=1}^p\theta_s T_s\). The fixed elements of the
pattern are in \(T_0\), which has zeroes for the free elements, while
the \(T_s\) with \(s>0\) all have zeroes for the fixed elements. In this
linear/quadratic case we have \begin{subequations}
\begin{align}
\mathcal{D}_sA&=T_s'CT+T'CT_s,\\
\mathcal{D}_{st}A&=T_s'CT_t+T_t'CT_s,
\end{align}
\end{subequations} which are symmetric matrices if order \(p\). Thus
\begin{equation}
\mathcal{D}_s\lambda_\nu=x_\nu'Q_sx_\nu,\label{eq-mdfadl}
\end{equation} where \(Q_s:=\mathcal{D}_sA=T_s'CT+T'CT_s\). It follows
that \begin{equation}
\mathcal{D}_{st}\lambda_\nu=2x_\nu'Q_s(\mathcal{D}_sx_t)+x_\nu'(\mathcal{D}_tQ_s)x_\nu.
\end{equation} From \eqref{eq-singvec} \begin{equation}
\mathcal{D}_tx_\nu=-(A-\lambda_\nu I)^+Q_tx_\nu,
\end{equation} and thus, or directly from \eqref{eq-hessvalsimple},
\begin{equation}
\mathcal{D}_{st}\lambda_\nu=-2x_\nu'Q_s(A-\lambda_\nu I)^+Q_tx_\nu+2x_\nu'T_s'CT_tx_\nu.\label{eq-mdfaddl}
\end{equation} Formulae \eqref{eq-mdfadl} and \eqref{eq-mdfaddl} apply
to all EVD problems where \(A=T'CT\) and \(T\) is a weighted linear
combination of known matrices, with or without intercept.

We now specialize to the problem of differentiating the trace norm
\begin{equation}
\sigma(T):=\|XT\|_\tau=\sum_{\nu=1}^m\lambda_\nu^\frac12(T'CT),
\end{equation} for which we have \begin{equation}
\mathcal{D}_s\sigma=\sum_{\nu=1}^m\lambda_\nu^{-\frac12}x_\nu'T'CT_sx_\nu=\text{tr}\ (T'CT)^{-\frac12}T'CT_s,
\label{eq-dsigma}\end{equation} where \((T'CT)^{-\frac12}\) is the
square root of the Moore-Penrose inverse of \(T'CT\). Next differentiate
\eqref{eq-dsigma} to find \begin{equation}
\mathcal{D}_{st}\sigma=-\frac14\sum_{\nu=1}^m\lambda_\nu^{-\frac32}(\mathcal{D}_s\lambda_\nu)(\mathcal{D}_t\lambda_\nu)+\frac12\sum_{\nu=1}^m\lambda_\nu^{-\frac12}(\mathcal{D}_{st}\lambda_\nu),
\end{equation} and we can substitute from formulae \eqref{eq-mdfadl} and
\eqref{eq-mdfaddl} for a final result.

If the \(T_s\) are unit matrices we find \begin{equation}
\mathcal{D}\sigma=CT(T'CT)^{-\frac12},
\end{equation} where the fixed elements correspond with zeroes in
\(\mathcal{D}\sigma\).

The repository has the file mdfa.R, which has functions (and numDeriv
checkers) to compute the first and second derivatives of the eigenvalues
of the general linear/quadratic problem and the first and second
derivatives of the trace norm for the MDFA problem.

\sectionbreak

\section*{References}\label{references}
\addcontentsline{toc}{section}{References}

\phantomsection\label{refs}

\end{document}